\documentclass[12pt,superscriptaddress,aps,prd,preprint]{revtex4}
\usepackage{graphicx}
\usepackage{amsmath}
\usepackage{subfigure}
\usepackage{slashed}

\def\slash#1{#1\!\!\! /}

\usepackage[latin1]{inputenc}

\begin{document}

\title{One-loop corrections in the $z=3$ Lifshitz extension of QED}

\author{M. Gomes}
\affiliation{Instituto de F\'\i sica, Universidade de S\~ao Paulo\\
Caixa Postal 66318, 05315-970, S\~ao Paulo, SP, Brazil}
\email{mgomes,fabriciomarques,ajsilva@if.usp.br}

\author{F. Marques}
\affiliation{Instituto de F\'\i sica, Universidade de S\~ao Paulo\\
Caixa Postal 66318, 05315-970, S\~ao Paulo, SP, Brazil}
\email{mgomes,fabriciomarques,ajsilva@if.usp.br}

\author{T. Mariz}
\affiliation{Instituto de F\'\i sica, Universidade Federal de Alagoas\\ 
57072-270, Macei\'o, Alagoas, Brazil}
\email{tmariz@fis.ufal.br}

\author{J. R. Nascimento}
\affiliation{Departamento de F\'{\i}sica, Universidade Federal da Para\'{\i}ba\\
Caixa Postal 5008, 58051-970, Jo\~ao Pessoa, Para\'{\i}ba, Brazil}
\email{jroberto,petrov@fisica.ufpb.br}

\author{A. Yu. Petrov}
\affiliation{Departamento de F\'{\i}sica, Universidade Federal da Para\'{\i}ba\\
Caixa Postal 5008, 58051-970, Jo\~ao Pessoa, Para\'{\i}ba, Brazil}
\email{jroberto,petrov@fisica.ufpb.br}

\author{A. J. da Silva}
\affiliation{Instituto de F\'\i sica, Universidade de S\~ao Paulo\\
Caixa Postal 66318, 05315-970, S\~ao Paulo, SP, Brazil}
\email{mgomes,fabriciomarques,ajsilva@if.usp.br}

\date{\today}

\begin{abstract}
In this work we study a $z=3$ Horava-Lifshitz-like extension of QED in (3+1) dimensions. We calculate the one-loop radiative corrections to the two and three-point functions of the gauge and fermion fields. Such corrections were achieved using the perturbative approach and a dimensional regularization was performed only in the spatial sector.
 Renormalization was required to eliminate the divergent contributions emergent from the photon and electron self-energies and from the three-point function. We verify that the one-loop vertex functions
 satisfy the usual Ward identities and using renormalization group methods we show that the model is asymptotically free.
\end{abstract}

\maketitle

\section{Introduction}

The possibility of Lorentz symmetry breaking, which began to be discussed in early 90s \cite{Kostel}, motivated  interest in the idea of space-time anisotropy based on the suggestion that spatial and temporal coordinates enter field theories in distinct ways; the resulting theory involves different orders in the derivatives with respect to time and space coordinates. Certainly, one of the initial motivations for this emerged from studies of condensed matter \cite{Lifshitz}, where this anisotropy was introduced for the development of effective models describing Lifshitz phase transitions. 

In the realm of relativistic quantum field theories, the first studies on anisotropic models were performed in \cite{Anselmi} where the renormalizability of scalar field theories with space-time anisotropy was discussed in great details. The interest on these theories was further enhanced after the famous Horava's paper \cite{PH}, where  models of this kind  were proposed within the gravity context. In that paper, and many others subsequently written, it was assumed that the theory is invariant under the rescaling $t\to b^{-z}t$, $x_i\to b^{-1}x_i$, with $z$ being a number called the critical exponent.  Clearly, for $z>1$ the order in spatial derivatives increases, which naturally can improve the renormalization properties of the theory. No difficulty with unitarity is to be expected since 
the canonical structure is preserved.
 It was argued in \cite{PH} that since for the critical exponent $z=3$ the gravitational coupling constant is dimensionless, it is natural to expect the power-counting renormalizability for the gravity with this value. Afterward, the theories with space-time anisotropy began to be referred as Horava-Lifshitz-like theories. Numerous issues related to different aspects of the Horava-Lifshitz (HL) gravity including exact solutions, algebraic structure, cosmological implications, etc. have been studied (for a review on Horava-Lifshitz gravity, see f.e. \cite{Visser}). At the same time, it is clear that the HL-like approach not only to gravity, but also to other field theories, could give interesting results because of the possibility of improving the renormalization behavior of the corresponding theories. 

The most important theory whose HL-like formulation must be considered is QED.
Recently, some important results were obtained for it in the case $z=2$, such as the one-loop corrections to the two-point function of the gauge field. There it was verified, that the dynamical restoration of the Lorentz symmetry occurs at low energies, at least in certain cases \cite{TM1}. The effective potential was already studied in \cite{petrov1}, its finite temperature generalization was obtained in \cite{petrov2}, and its gauge dependence in \cite{petrov3}. At the same time, the consideration of the case with odd $z>1$ seems to be very important. This is confirmed by the fact that in the paper \cite{GN}, where the HL-like analogue of the Gross-Neveu model was considered, it was shown that odd values of $z$ are more interesting, allowing for nontrivial mass generation. Therefore, the study of quantum dynamics of theories with generic odd $z$ seems to be more interesting since it allows for straightforward comparison with the Lorentz-invariant case. Up to now, the  main results in this direction were presented in \cite{Taiuti}, where some aspects of HL-like QED with $z=3$ were analyzed, in \cite{Bakas}, where the triangle anomaly  was studied, and in \cite{Dhar} where the $z=3$ four-fermion theory has been considered.
Therefore, it is relevant to pursue the study of  perturbative aspects of the HL-like QED with $z=3$ in more details. 

In this work we study the possibility of symmetry restoration in a $z=3$ Lifshitz extension of QED in (3+1) dimensions, through the calculation of the one-loop radiative corrections. However, before entering into this analysis we would like to make a comment on a peculiarity of the regularization method we use when  applied to a model with a  generic critical exponent. 
To keep the ultraviolet divergences under control, we employ dimensional regularization in the spatial part of the 
Feynman integrals. Besides that, to extract its low energy behavior we examine the  Taylor expansions of  the integrands  around zero external momenta. 
Proceeding in  this way, we found that the divergences manifest themselves as poles at values of  the degree of superficial divergence multiples of $2 z$. Therefore, for many superficially divergent diagrams  the dimensional regularization gives finite results. At first sight in our case,   we would conclude that there is the emergence of a finite Maxwell  term. However, this is a hasty conclusion  as the  radiative correction to the self energy of the spinor field turns out to be divergent, requires
a counterterm and, as a consequence, generates divergences also in the coefficient of the Maxwell term.

This work is organized as follows. In  section 2 we present the model, in section 3 we perform the calculations of the one-loop corrections to quadratic actions of gauge and spinor fields and to the three-point vector-spinor vertex function, in section 4, we discuss the Ward identities, and in section 5 we employ renormalization group methods to investigate the behavior of the model both  at high and low energies. We summarize our results in the last section. Throughout this work, we use the Minkowski metric $g_{\mu\nu}=diag(1,-1,-1,-1)$.

\section{The Model}

The Lagrangian describing the $z=3$ HL-like QED in $(3+1)$ dimensions looks like
\begin{eqnarray}
\mathcal{L}&=&-\frac{1}{2}F_{0i}F^{0i}-\frac{a_{3}^{2}}{4}F_{ij}\Delta^2F^{ij}+
\nonumber\\ &+& \bar\psi\left[i\gamma^0(\partial_0-ieA_0)+b_1(i\gamma^i(\partial_i-ieA_i))+b_3(i\gamma^i(\partial_i-ieA_i))^3-m^3\right]\psi.
\end{eqnarray}
where $\Delta=\partial_{i}\partial^{i}$. The corresponding action possesses  an Abelian gauge symmetry with the following transformations:
\begin{equation}
\psi\rightarrow e^{ie\alpha}\psi,\hspace{0.5cm}\bar\psi\rightarrow e^{-ie\alpha}\bar\psi,\hspace{0.5cm}A_{0,i}\rightarrow A_{0,i}+\partial_{0,i}\alpha,
\end{equation}
and the mass dimensions are
\begin{eqnarray}
[A_i]&=& 0,\hspace{1cm}[A_0]=2,\hspace{1cm}[m]=[e]=1,\hspace{1cm}[\psi]=\frac{3}{2},\nonumber\\
&&[a_{3}] = [b_{3}]=0,\hspace{1cm}[b_{1}]=2.
\end{eqnarray}

 In order to obtain the photon propagator and keep it strictly diagonal, the following gauge fixing will be used (cf. \cite{TM1}):
\begin{equation}
\mathcal{L}_{gf}=-\frac{1}{2}\left[ (a_{3}\Delta)^{-1}\partial_0A^0+a_{3}\Delta\partial_iA^i\right] ^2.
\end{equation}
The resulting Feynman rules  are given by the list below (here and henceforth we  follow the notation $\slash k\equiv k_i \gamma_i$ and $k^2\equiv \vert \vec k\vert^2$)

\begin{eqnarray}
\raisebox{-0.0cm}{\includegraphics[angle=0,scale=0.8]{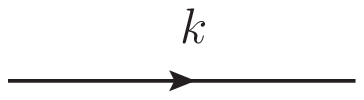}}
&& S(k)=\frac{i}{\gamma^0k_0+b_{3}\slash{k} k^2-m^3}=i\frac{\gamma^0k_0+b_{3}\slash{k}k^2+m^3}{k_0^2-b_{3}^{2}k^6-m^6};\\
\raisebox{-0.2cm}{\includegraphics[angle=0,scale=0.8]{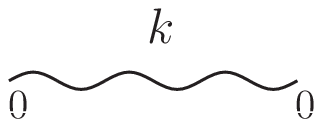}}
&&D_{00}(k)=-\frac{ia_{3}^{2}k^4}{k_0^2-a_{3}^{2}k^6};\\
\raisebox{-0.4cm}{\includegraphics[angle=0,scale=0.8]{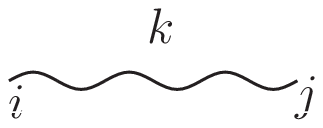}}
 && D_{ij}(k)=-\frac{ig_{ij}}{k_0^2-a_{3}^{2}k^6};\\[1pt]
 \raisebox{-0.5cm}{\includegraphics[angle=0,scale=0.8]{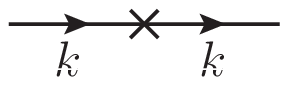}}
 && V_{2}(k,p)= -b_{1}\slash{k};\\[10pt]
  \raisebox{-0.7cm}{\includegraphics[angle=0,scale=0.8]{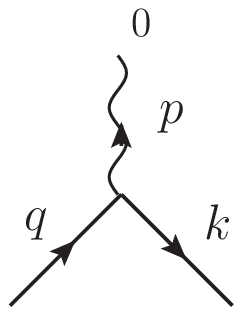}}
 &&V_3^0(k,p)=e(2\pi)^4\delta^4(p+k-q)\gamma^{0};
 \end{eqnarray}
\begin{eqnarray}
\raisebox{-0.6cm}{\includegraphics[angle=0,scale=0.8]{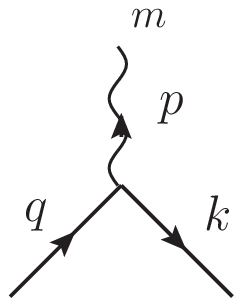}}
 \nonumber&&V_3^{m}(k,p)=V_3^{(1)m}+V_3^{(2)m};\nonumber\\
 V_{3m}^{(1)}&=&eb_1(2\pi)^4\delta^4(p+k-q)\gamma_m;\nonumber\\
V_{3m}^{(2)}&=&eb_3(2\pi)^4\delta^4(p+k-q)\left(-\gamma_m (\vec p+\vec{k})^2+\slash{p}\gamma_m \slash{k}-2\slash{k} k_m\right).
\end{eqnarray}
\begin{eqnarray} 
\raisebox{-.3cm}{\includegraphics[angle=0,scale=0.8]{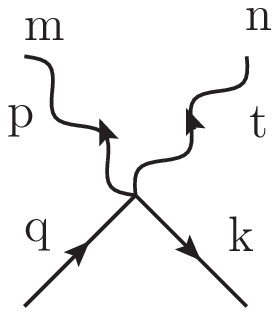}}
 \nonumber&& V_{4mn}(p,k,t)=e^2b_3(2\pi)^4\delta^4(p+k+t-q)\left(-2g_{mn}(p\!\!\!/+t\!\!\!/)-4g_{mn}k\!\!\!/
 \right.\nonumber\\ &&\left. -\gamma_m \slash t\gamma_n -\gamma_n\slash p\gamma_m -\gamma_m\slash k\gamma_n -\gamma_n\slash k\gamma_m \right).\label{10}
\end{eqnarray}
There is also the quintuple vertex
\begin{eqnarray}
V_5^{mnl}(p,k,t,q)=    e^3b_3(2\pi)^4\delta^4(p+k+t+q)\gamma^m\gamma^n\gamma^l.
\end{eqnarray}
% However, it is irrelevant within calculations of the one-loop two-point functions, hence, within our studies we disregard it.

We can calculate the superficial degree of divergence  for a generic graph $\gamma$.  It is given by
\begin{equation}
\delta(\gamma)=6-2E_0-\frac{3}{2}E_{\psi}-2V_2-V_3^0-3V_3^{(1)}-V_3^{(2)}-2V_4-3V_5,
\end{equation}
where $E_0$ is the number of external $A_0$ lines (we note that $\delta(\gamma)$ does not depend explicitly on the number of external $A_i$ lines since the $A_i$ has zero mass dimension), $E_{\psi}$ is the number of external spinor legs, $V_{2}$, $V_3^0$, $V_3^{(1)}$, $V_3^{(2)}$, $V_4$ and $V_5$ are the numbers of the corresponding vertices. The model is  super renormalizable; thus, in spite of being marginal, four fermion self interactions are not generated to any finite order of perturbation.  Also, taking
in account gauge invariance, we may verify that there are no divergent corrections to the
vertices $F_{ij}\Delta^2F^{ij}$, $\bar \psi[(i\gamma^i(\partial_i-ieA_i))^3]\psi$ and $\bar \psi[(i\gamma^i(\partial_i-ieA_i))^2]\psi$. For simplicity, in what follows  we take $b_{3}=a_{3}$.
Furthermore, also  due  to gauge invariance, the two point function of the $F_{0i}$ field, $\langle F_{0i}F^{0i}\rangle$, turns out to be  finite. 

 By adopting dimensional regularization in the spatial part of the relevant Feynman integrals, we may verify that a typical integral to be determined, has the form
\begin{equation}
K(x,y,w)=\int \frac{dk_{0} d^{d}k}{(2\pi)^{d+1}} \frac{k_{0}^{2x}k^{y}}{(k_{0}^{2}-a_{3}^{2}k^{6}-m^{6})^w}
\end{equation}
and is given by
\begin{eqnarray}
K(x,y,w)&=&\frac{1}{3 \Gamma \left(\frac{d}{2}\right) \Gamma (w)}
2^{-d-1} \pi ^{-\frac{d}{2}-1} i^{x+1} (-1)^{w+x} a_{3}^{\frac{1}{3} (-d-y)} m^\delta \times\nonumber\\
   &\times&\Gamma
   \left(\frac{2x+1}{2}\right) \Gamma
   \left(\frac{d+y}{6}\right) \Gamma \left(-\frac{\delta}{6} \right).
\end{eqnarray}

\noindent
We see that divergences will appear only if its degree of superficial divergence $\delta= 3+d+y+6x-6w=0,6,12,\ldots$. Thus, by employing dimensional regularization in the spatial part, the two point function
$\langle F_{ij} F^{ij}\rangle$ will require counterterms of the type  $a_{1}F_{ij}F_{ij}$ and $a_{2}F_{ij}\Delta F_{ij}$ to become finite. Similarly, in the matter sector the coupling $b_{1}$ will have to be adjusted to secure the finiteness through the calculations. These are the only divergences to be found in this model.

As a consequence of the above remarks, we may write down the complete bare Lagrangian as
\begin{eqnarray}
\mathcal{L}&=&-\frac{1}{2}F_{0i}F^{0i}-\frac{a_{1}}{4}F_{ij}F^{ij}-\frac{a_{2}}{4}F_{ij}\Delta F^{ij}-\frac{a_{3}^{2}}{4}F_{ij}\Delta^2F^{ij}
\nonumber\\ &+& \bar\psi\left[i\gamma^0(\partial_0-ieA_0)+b_1(i\gamma^i(\partial_i-ieA_i))+b_3(i\gamma^i(\partial_i-ieA_i))^3-m^3\right]\psi
\end{eqnarray}
and define its renormalized version by performing the reparametrization: $A_{0}= Z^{1/2}A_{0r}$, $A_{i}=Z^{1/2}A_{ir}$, $\psi=Z^{1/2}_{2}\psi_{r}$; $a_{i} = \frac{Z_{a_{i}}}{Z} a_{ir}$ for $i=1,2$ and $ b_{1}=\frac{Z_{b_{1}}}{Z_{2}} b_{1r}$. The Lagrangian becomes
\begin{eqnarray}
\mathcal{L}&=&-\frac{Z}{2}F_{0i}F^{0i}-\frac{a_{1r}Z_{a_1}}{4}F_{ij}F^{ij}-\frac{a_{2r}Z_{a_{2}}}{4}F_{ij}\Delta F^{ij}-\frac{a_{3}^{2}Z}{4}F_{ij}\Delta^2F^{ij}+ Z_{2}\bar\psi \times 
\\ &&\left[  i\gamma^0(\partial_0-ieZ^{1/2}A_0)+\frac{Z_{b_{1}} b_{1r}}{Z_{2}}(i\gamma^i(\partial_i-ieZ^{1/2}A_i))+b_3(i\gamma^i(\partial_i-ieZ^{1/2}A_i))^3-m^3\right]\psi ,\nonumber
\end{eqnarray}
where now all basic fields are the renormalized ones.
In the next section we will examine the one-loop contributions to the renormalization factors; for simplicity, the parameters $a_{1}, a_{2}$ and $b_{1}$ will be treated perturbatively so that they will not contribute to the free propagators of the basic fields. As we shall verify, the wave function renormalizations $Z$ and $Z_{2}$ turn out to be finite.

\section{One-loop corrections}

\subsection{The pure gauge sector}

\begin{figure}[ht]
%\raisebox{-0.8cm}{\includegraphics[angle=0,scale=1.3]{12a.eps}}
\includegraphics[width=0.37\columnwidth]{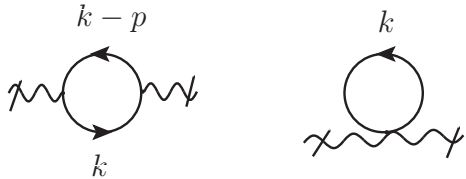}
\caption{Radiative corrections to the photon two point function.\label{Figura2}}
\end{figure}

We begin our analysis of the radiative corrections by calculating the contributions to the two-point function of the gauge field coming from  the graphs depicted in 
Fig. \ref{Figura2}. 

%\begin{eqnarray}
%\raisebox{-0.8cm}{\includegraphics[angle=0,scale=1.3]{2.eps}}
 %&=&(i\Pi_1^{00}(p),i\Pi_2^{0i}(p),i\Pi_3^{ij}(p))\\
 %\nonumber\\
%\raisebox{-0.5cm}{\includegraphics[angle=0,scale=1.0]{10.eps}}
% \,\,\,\,\,\,\,\,&=&i\Pi^{ij}_4(p)
%\end{eqnarray}

Our first aim is to obtain a correction to the (generalized) Maxwell term. The tadpole, the second graph in Fig. \ref{Figura2}, does not contribute to the Maxwell term because it is based on the vertex $V_4^{mn}(p,k,t)$ which contains only one power of the external momentum, and therefore  cannot generate  terms quadratic in $F_{\mu\nu}$. This diagram only serves to cancel non-gauge-invariant contributions from the other graphs.

The photon self-energy gives us three contributions coming from the first graph in Fig. \ref{Figura2}, $i\Pi_{00}(p)$, with two external $A_0$ fields, $i\Pi_{0i}(p)$, with one $A_i$ and one $A_0$ and $i\Pi_{ij}$,with two $A_i$ external fields. Up to first order in $b_{1}$, each  of these corrections receives two contributions. We will examine separately each of them.

1. Contributions to the two point function $\langle A_{0}A_{0}\rangle$. 
\begin{equation}
\Pi^{00}(p)=\Pi^{00}_{1}(p)+\Pi^{00}_{2}(p)
\end{equation}
where $\Pi^{00}_{1}$ and $\Pi^{00}_{2}$ are, respectively, the terms of zero and first order in $b_{1}$:
\begin{equation}
\Pi^{00}_{1}(p)=e^2\mu^{3-d}\int\frac{dk_0d^dk}{(2\pi)^{d+1}}{\rm tr}[\gamma^0S(k)\gamma^0S(k-p)],\label{11}
\end{equation}
which is independent of $b_{1}$, and $\Pi^{00}_{2}$ which collects the result of one insertion of the bilinear part of the $b_{1}$ vertex,
\begin{equation}
 \Pi^{00}_{2}(p)=-ie^2b_{1}\mu^{3-d}\int\frac{dk_0d^dk}{(2\pi)^{d+1}}{\rm tr}[\slash{k}S(k)\gamma_{0}S(k - p)\gamma_{0} S( k)+S(k) \gamma_{0} S(k - p) (k\!\!\!/-p\!\!\!/) S(k - p) \gamma_{0}],\label{12}
\end{equation}
 where, as usual, the parameter $\mu$ with mass dimension one was introduced to fix the dimension of the integral  to its value in three  spatial dimensions.
 In order to carry out the above integration, we perform a derivative expansion in the external momentum,
\begin{equation}
 \Pi^{00}(p)=\Pi^{00}(0)+ \Bigl.\frac{p_i p_j}{2}\frac{\partial^2\Pi^{00}}{\partial p_{i}\partial p_{j}}\Bigr \vert_{p=0} + \cdots,
 \end{equation} 
   and use integration over $d$-dimensional spherical coordinates. From now on, $\Pi^{00}(p)$ will represent the parts of (\ref{11}) and (\ref{12}) non-trivially contributing to $F_{0i}F^{0i}$, i.e., $\Pi^{00}(p)$ will contain only terms with two spatial external momenta. After evaluating the trace, disregarding the terms which are odd with respect to momenta, we obtain

\begin{eqnarray}
 \nonumber \Pi^{00}_{1}(p)&=&-e^2\mu^{3-d}\frac{2 \pi ^{d/2}}{\Gamma \left(\frac{d}{2}\right)}\,\,\vec{p}^2\int\frac{dk_0dk}{(2\pi)^{d+1}}k^{d-1}\left[\frac{216k_0^2 k^{10}}{d}+\frac{32k_0^4 k^4 }{d}-\frac{32  m^{12} k^4 }{d}+\frac{8m^6 k^{10}}{d}\right.\\
 &&\left.+\frac{40k^{16} }{d}+4k_0^4 k^4 -4m^{12} k^4 -8m^6 k^{10} -4k^{16} \right]\frac{1}{\left(k_0^2-a_{3}^{2}k^6-m^6\right)^4}.
\end{eqnarray}
 Finally, after  integrating, we obtain
\begin{equation}
\Pi_{1}^{00}(p)=-i\frac{e^2a_{3}^{\frac{2-d}{3}}\mu^{3-d}   m^{d-5} \Gamma \left(\frac{5}{6}-\frac{d}{6}\right) \Gamma \left(\frac{d+4}{6}\right)}{3 (2\pi)^{\frac12(d+1)}\Gamma \left(\frac{d}{2}+1\right)}\vec{p}^2.
\end{equation}

Analogously, we have
\begin{equation}
\Pi^{00}_{2}(p)=
-i\frac{e^2 b_{1} a_{3}^{\frac{1}{3}-\frac{d}{3}}\mu^{3-d} 
   m^{d-7} \Gamma
   \left(\frac{7}{6}-\frac{d}{6}\right) \Gamma \left(\frac{d+8}{6}\right)
   }{3 (2 \pi )^{\frac{1}{2} (d+1)}  \Gamma \left(\frac{d}{2}+1\right)} {\vec p}^{2}.
 \end{equation}
 
 Therefore,
 \begin{equation}
 \Pi^{00}= -i\alpha_{1}\vec{p}^{2},
 \end{equation}
with
\begin{equation}
\label{alpha1}
\alpha_{1}=\frac{e^2a_{3}^{\frac{2-d}{3}}\mu^{3-d}   m^{d-5} \Gamma \left(\frac{5}{6}-\frac{d}{6}\right) \Gamma \left(\frac{d+4}{6}\right)}{3 (2\pi)^{\frac12(d+1)}\Gamma \left(\frac{d}{2}+1\right)}+\frac{e^2 b_{1} a_{3}^{\frac{1}{3}-\frac{d}{3}}\mu^{3-d} 
   m^{d-7} \Gamma
   \left(\frac{7}{6}-\frac{d}{6}\right) \Gamma \left(\frac{d+8}{6}\right)
  }{3 (2 \pi )^{\frac{1}{2} (d+1)}  \Gamma \left(\frac{d}{2}+1\right)}.
\end{equation}
2. Using a similar procedure, the quantities $\Pi_{0i}(p)$ and $\Pi_{ij}(p)$, that from now on will  represent   contributions to $F_{0i}F^{0i}$ and $F_{ij}F^{ij}$, can be found. They are
  
\begin{eqnarray}
\Pi_{0i}(p)&=&e\mu^{3-d}\int\frac{dk_0d^dk}{(2\pi)^{d+1}}{\rm tr}\left\{V_{3i}(k,-p)S(k)\gamma^0S(k-p)-i b_{1}V_{3j}^{(2)}(k,-p)\times\right.\nonumber\\&\times&\left.[S(k)\slash{k} S(k)
 \gamma_{0}S(k-p)+ S(k) \gamma_{0}S(k-p)(\slash{k}-\slash{p})S(k-p)]\right\},
\end{eqnarray}
\begin{eqnarray}
\label{13}
\Pi_{ij}(p)&=& \mu^{3-d}\int\frac{dk_0d^dk}{(2\pi)^{d+1}}{\rm tr}\left\{V_{3i}(k,-p)S(k)V_{3j}(k-p,p)S(k-p)] -\right.\nonumber\\ &-&\left.
i b_{1}V_{3i}^{(2)}(k,- p)%
[ S(k)\slash{k} 
 S(k)V_{3j}^{(2)}(k-p,p) S(k-p)-\right.\nonumber\\ &-&\left.
 S(k) V_{3j}^{(2)}(k-p,p)S(k-p)(\slash{k}-\slash{p})S(k-p)]\right\},
\end{eqnarray}
which yields
\begin{equation}
\Pi_{0i}(p)= -i \alpha_{1}p_{0}p_{i},
\end{equation}
and
\begin{equation}
\Pi_{ij}=\Pi_{1ij} +\Pi_{2ij},
\end{equation}
with
\begin{eqnarray}
\Pi_{1ij}&=&-i\alpha_{1}p_{0}^{2}\delta_{ij},\\
\Pi_{2ij}&=&i\alpha_{2} (p_{i}p_{j}-{\vec{p}}^2\delta_{ij}),
\end{eqnarray}
where $\alpha_{2}$ is not finite for $d=3$ and is given by
\begin{eqnarray}
\alpha_{2}&=&-\Big[\frac{  e^2 a_{3}^{\frac{4}{3}-\frac{d}{3}} \mu^{3-d} m^{d-1} d(d-2) 
   \Gamma \left(\frac{1}{6}-\frac{d}{6}\right) \Gamma
   \left(\frac{d+2}{6}\right) }{12 (2\pi)^{\frac12(d+1)} \Gamma
   \left(\frac{d}{2}+1\right)}+\nonumber\\
   &&+\frac{b_{1} e^{2}a_{3}^{1-\frac{d}{3}}\mu^{3-d}m^{d-3}(d-2)(d-4)\Gamma\left(\frac{1}{2}-\frac{d}{6}\right)\Gamma\left(\frac{d}{6}+1\right)}{6  (2 \pi )^{\frac{1}{2} (d+1)}\Gamma\left(\frac{d}{2}+1\right)}\Big].\label{14}
\end{eqnarray}

Observe that the Ward identities
\begin{equation}
\label{ward1}
p^{0}\Pi_{00}+p^{i}\Pi_{i0}=0
\end{equation}
and
\begin{equation}
\label{ward2}
p^{0}\Pi_{0j}+p^{i}\Pi_{ij}=0
\end{equation}
are automatically satisfied. Notice also that  the coefficients associated with $\Pi_{00}$, $\Pi_{0i}$ and  $\Pi_{1ij}$  are the same, and therefore it is possible to simplify the sum of these corrections as
\begin{equation}
\Pi^{00}(p)A_0(p)A_0(-p)+\Pi^{0i}(p)A_0(p)A_i(-p)+\Pi_{1ij}(p)A_i(p)A_j(-p)=\alpha_1 F_{0i}F_{0i}.
\end{equation}

 As   the gauge structure  in the above result is preserved with $\alpha_{1}$  finite at $d=3$, the wave function renormalization for the the gauge fields, $Z$, may be taken equal to one.
Moreover, as we have anticipated, at $d=3$, $\Pi_{2ij}$ presents a divergence to first order in $b_{1}$. The cancellation of  this divergence  is accomplished by writing  $Z_{a_{1}}=1+\delta Z_{a_{1}}$  with
\begin{equation}
a_{1r}\delta Z_{a_{1}} =  \frac{b_{1}e^{2}}{6 \pi^{2}(d-3)},
\end{equation} 
 that effectively removes the pole part of (\ref{14}).

In this pure gauge sector, there is also a divergence of $\Pi_{ij}$ containing four factors of the momentum. To get its explicit value,  we need  to calculate the fourth derivative of  (\ref{13}) with respect to the external momentum and with $b_{1}=0$. We obtain the result
\begin{eqnarray}
\Pi_{ij}&=& i \alpha_{3}(p_{i}p_{j}
   -{\vec p}^2 \delta_{ij}){\vec p}^2,\\
\alpha_{3}&=&\frac{ e^2 a_{3}^{2
   \left(1-\frac{d}{6}\right)} 2^{\frac{5}{2}-\frac{d}{2}} \pi ^{-\frac{d}{2}-\frac{1}{2}}   m^{d-3}\mu^{3-d} \Gamma
   \left(\frac{1}{2}-\frac{d}{6}\right) \Gamma \left(\frac{d}{6}\right) }{5 \Gamma \left(\frac{d}{2}-2\right)}\label{a2}\\
&=&\frac{6e^2a_{3}}{5\pi^{2}(d-3)}+\mbox{Finite term}.\nonumber
\end{eqnarray}
To cancel the above divergence we adjust the renormalization of the parameter $a_{2}$ so that $Z_{a_{2}}=1+ \delta Z_{a_{2}}$, with
 \begin{equation}
a_{2r} \delta Z_{a_{2}}= \frac{6a_{3}e^2}{5 \pi^2(d-3)}.
 \end{equation}
 %With these corrections calculated,  the pure gauge sector of the effective one-loop  Lagrangian can be written as follows:
With these corrections calculated, up to a term proportional to
$F_{ij} \triangle^2 F_{ij}$ that is not renormalized, the pure gauge sector of the effective one-loop Lagrangian can be written as follows:
\begin{eqnarray}
\mathcal{L}_{\gamma}=\frac{1}{2}(1+\alpha_1)F_{0i}F_{0i}-\frac{1}{4} (a_{1r}+ \alpha_{2Fin})F_{ij}F_{ij}-\frac{1}{4}(a_{2r}+ \alpha_{3Fin})F_{ij}\Delta F_{ij}-\frac{a^2_3}{4}F_{ij}\Delta^2 F_{ij}.
\end{eqnarray} 
where $\alpha_{2Fin}= \alpha_{2}-\frac{b_{1}e^{2}}{6 \pi^{2}(d-3)}$ and $\alpha_{3Fin}=\alpha_{3}-\frac{6a_{3}e^2}{5 \pi^2(d-3)}$ are,  respectively, the finite parts of (\ref{14})
 and (\ref{a2}).
The new terms  with coefficients  containing $a_{1r}$ or $a_{2r}$  do not induce new divergences and can be naturally treated as  small perturbations to the classical action. For simplicity, they will be omitted henceforth.
Notice also that, due to gauge invariance, graphs with six external spatial gauge field lines
although individually logarithmically divergent provide a finite result. 

\subsection{Corrections to the two point function of the spinor field}

\begin{figure}[ht]
\raisebox{-0.5cm}{\includegraphics[width=0.17\columnwidth]{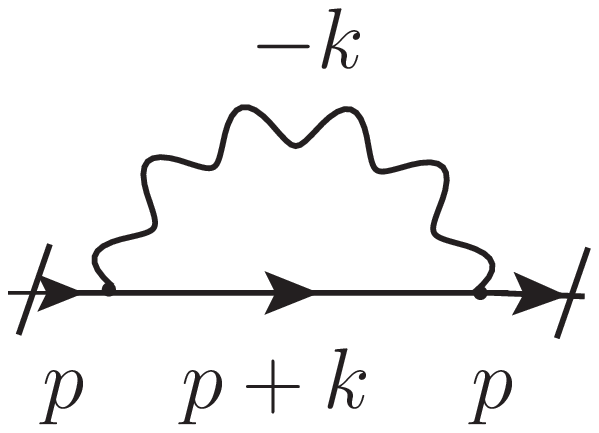}}\hspace{4.0cm} 
\raisebox{-0.5cm}{\includegraphics[width=0.12\columnwidth]{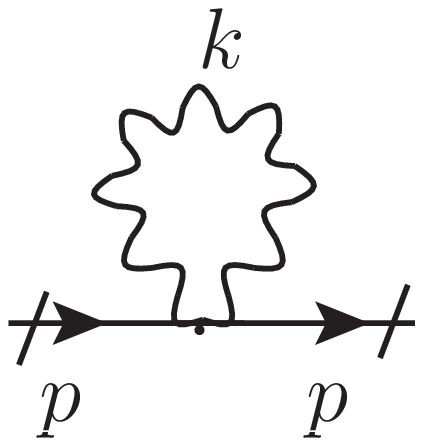}}
\caption{One-loop self energy graphs. There are two contributions for the first graph: one in which the wavy line corresponds to the propagator of the $A_{0}$ field and another
where it represents the propagatorfor the $A_{i}$ field. \label{fig10} }
\end{figure}

 The next step is to calculate the one-loop corrections for the spinor sector of the QED.  We have two graphs, which are, as in the previous case, the self-energy and  the tadpole. We will not consider the modifications in these contributions due to insertions of the vertices $V_{2}$ or $V_{3}^{(1)}$ as they are finite and, after renormalization, $b_{1}$ will be taken very small.  For the fermion self-energy, $\Sigma(p)$, we have two contributions, one with two temporal vertices and one with two spatial vertices. There are no mixed contributions because our photon propagator is strictly diagonal. These contributions, which correspond to the first graph in Fig. \ref{fig10} are given by the expressions below:
\begin{eqnarray}
%\raisebox{-0.5cm}{\includegraphics[angle=0,scale=1.0]{fig10.eps}}
 i\Sigma(p)=i\Sigma_{1}(p)+i\Sigma_{2}(p)
\end{eqnarray}

\begin{eqnarray}
\label{26}
i\Sigma_{1}(p)&=&-e^2\mu^{3-d}\int\frac{dk_0d^dk}{(2\pi)^{d+1}}\gamma^0S(p+k)\gamma^0D_{00}(-k),\\
\label{27}
i\Sigma_{2}(p)&=&-\mu^{3-d}\int\frac{dk_0d^dk}{(2\pi)^{d+1}}V_{3i}(p+k,-k)S(p+k)V_{3j}(p,-k)D_{ij}(-k).
\end{eqnarray}

From now on $i\Sigma_{1}(p)$  will be the part of (\ref{26}) contributing to Dirac-like correction. Thus we must consider only terms up to the linear order in $p_0$ and $p_i$. Hence $i\Sigma_{1}(p)$ can be written as 
\begin{eqnarray}
i\Sigma_{1}(p)&=&-e^2\mu^{3-d}\frac{2 \pi ^{d/2}}{\Gamma \left(\frac{d}{2}\right)}\int\frac{dk_0dk}{(2\pi)^{d+1}}\frac{m^3k^{d+3}}{(k_0^2-a_{3}^{2}k^6-m^6)(k_0^2-a_{3}^{2}k^6)}+\\
&+&e^2\frac{2 \pi ^{d/2}}{\Gamma \left(\frac{d}{2}\right)}\int\frac{dk_0dk}{(2\pi)^{d+1}}\frac{k^{d-1}}{\left(k_0^2-a_{3}^{2}k^6-m^6\right)^2\left(k_0^2-a_{3}^{2}k^6\right)}\left[p_{i}\gamma^{i}\left(-\frac{2a_{3}k_0^2k^6}{d}+\frac{2a_{3} m^6 k^6}{d}\right.\right.\nonumber\\ &-&\left.\left.\frac{4 a_{3}^{3}k^{12}}{d}-a_{3}k_0^2  k^6+a_{3}m^6  k^6+a_{3}^{3} k^{12}\right)
+4 a_{3}^{2}\gamma^0 p_0 k_0^2 k^4\right],\nonumber
\end{eqnarray}
and, after the integration, we have
\begin{eqnarray}
\Sigma_{1}(p)&=&\frac{e^2\mu^{3-d} 2^{-d-1} \pi ^{-\frac{d}{2}-\frac{1}{2}} m^{d-3}}{9 \Gamma \left(\frac{d}{2}\right)} \left[-3 a_{3}^{(2-d)/3} m^{-2} \gamma ^0 p_0 \Gamma \left(\frac{d+4}{6}\right) \Gamma \left(-\frac{d}{6}-\frac{7}{6}\right)\right.\label{sigma1}\\
&&\left.+a_{3}^{1-d/3} p_i\gamma^i \Gamma \left(\frac{d}{6}+1\right) \Gamma \left(-\frac{d}{6}-\frac{3}{2}\right)+3 a_{3}^{(2-d)/3} m \Gamma \left(-\frac{d}{6}-\frac{1}{6}\right) \Gamma \left(\frac{d+4}{6}\right)\right].\nonumber
\end{eqnarray}

Observe that the term proportional to $p_i\gamma^i $ in the above expression diverges at $d=3$ with a pole term
\begin{equation}
- \frac{e^2}{48\pi^2}\frac{\gamma^i p_i}{d-3}.
\end{equation}

Using a similar procedure, we can find the Dirac-like contribution of $\Sigma_{2}(p)$, considering only terms that are linear in $p_0$ and $p_i$ and applying $d$-dimensional spherical coordinates to calculate the integrals, we get
\begin{eqnarray}
\nonumber\Sigma_{2}(p)&=&\frac{e^2\mu^{3-d} 2^{-d-2} \pi ^{\frac{1}{2} (-d-1)} m^{d-5} }{9 \Gamma \left(\frac{d}{2}\right)} \left[ a_{3}^{(2-d)/3}d \Gamma \left(\frac{d+4}{6}\right) \Gamma \left(-\frac{d}{6}-\frac{7}{6}\right)\right. \nonumber\\&\times& \left.\left(-(d+7) m^3+6d \gamma^0 p_0\right)+\right.\nonumber\\
&-&\left.\frac{2}d a_{3}^{1-d/3} m^2 p_i\gamma^i \{d[d (d+6)-12] +36\} \Gamma \left(\frac{d}{6}+1\right) \Gamma \left(-\frac{d}{6}-\frac{3}{2}\right)\right].\label{sigma2}
\end{eqnarray}
Here the pole part of $\Sigma_2$ is
\begin{equation}
-\frac{9 e^2}{16 \pi^2}\frac{\gamma^i p_i}{d-3}.
\end{equation}

Consequently, $\Sigma(p)=\Sigma_1(p)+\Sigma_2(p)$ is
\begin{eqnarray}
\Sigma(p)&=&\alpha_4 p_0\gamma^0+\alpha_5 p_i\gamma^i-\alpha_6 m^3,
\end{eqnarray}
where
\begin{eqnarray}
\label{alpha3}
\alpha_{4}&=& \frac{  2^{-d-1} \pi ^{-\frac{d}{2}-\frac{1}{2}} e^2 
   a_{3}^{\frac{2-d}{3}} \left( d-1\right) m^{d-5} \mu^{3-d}\Gamma
   \left(-\frac{d}{6}-\frac{7}{6}\right) \Gamma \left(\frac{d+4}{6}\right)
   }{3 \Gamma \left(\frac{d}{2}\right)} ,\\
   \label{alpha4}
\alpha_{5}&=&\frac{ 2^{-1-d} \pi ^{-\frac{d}{2}-\frac{1}{2}} e^2 a_{3}^{1-\frac{d}{3}}
   \left[d[d(d+6)-11]+36\right] m^{d-3} \mu^{3-d}\Gamma
   \left(-\frac{3}{2}-\frac{d}{6}\right) \Gamma \left(\frac{d}{6}+1\right)
   }{9d \Gamma \left(\frac{d}{2}\right)},\\
 \label{alpha5}
   \alpha_{6}&=&\frac{ 2^{-d} \pi ^{\frac{1}{2} (-d-1)} e^2 a_{3}^{\frac{2}{3}-\frac{d}{3}}
   m^{d-5}\mu^{3-d} \Gamma \left(\frac{5}{6}-\frac{d}{6}\right) \Gamma
   \left(\frac{d+4}{6}\right)}{\Gamma \left(\frac{d}{2}\right)}.
\end{eqnarray}

As we can see, the $\Sigma(p)$ (namely, $\alpha_5$)  diverges when we set $d=3$.  To be more precise,  as mentioned earlier, the divergence is present in the term $\bar{\psi}\gamma^ip_i\psi$,
\begin{equation}
\alpha_5=-\frac{7e^2}{12 \pi^2 (d-3) }+ \mbox{Finite terms},
\end{equation}
which may be absorbed in the parameter $b_{1}$.

The tadpole contribution may be obtained from the terms in the Lagrangian linear in the spatial derivative and containing two gauge fields. These two fields are then contracted and, to evade infrared divergences, we replace the gauge propagator by
\begin{equation}
-i \frac{g_{ij}}{k_{0}^{2}-a_{3}^{2}k^{6}-M^{6}},
\end{equation}
 the auxiliary mass parameter $M$ to be  eliminated at the end of the calculation. We then have

\begin{eqnarray}
 &&i\Sigma_{3}(p)=-i\mu^{3-d}(d+2) e^2 a_{3}\slash p \int\frac{dk_0d^dk}{(2\pi)^{d+1}}\frac{1}{k_{0}^{2}-a_{3}^{2}k^{6}-M^{6}},
\end{eqnarray}
corresponding to the second graph in Fig. \ref{fig10}.  After performing the integration, we get
 
\begin{eqnarray}
\label{sigma3}
\Sigma_{3}(p)&\equiv & \alpha_{7}\slash p=\frac{2^{-d-1} (d+2) \pi ^{-\frac{d}{2}-\frac{1}{2}} e^2 a_{3}^{1-\frac{d}{3}}
  \left(\frac{\mu}{M}\right)^{3-d}  \Gamma \left(\frac{1}{2}-\frac{d}{6}\right) \Gamma \left(\frac{d}{6}\right)
   \slash p}{3 \Gamma \left(\frac{d}{2}\right)}\nonumber \\&\approx& - \frac{5 e^2}{4 \pi^{2}(d-3)}\slash p- \frac{5 e^{2}}{4 \pi^{2}}\ln \frac{M}{\mu}\slash p,
\end{eqnarray}
where in the second line of the above regulated expression we also quoted both its pole part at $d=3$ and also the dependence on $M$ of its finite part. Notice that as $M$ tends to zero $\Sigma_{3}$ develops a logarithmic divergence. This divergence is however innocuous as it can be absorbed in the renormalization of the parameter $b_{1}$.
 
We may write now the one-loop corrected Dirac-like Lagrangian as
\begin{equation}
\mathcal{L}_{\bar\psi\psi}=\bar\psi\left[i\partial_0\gamma^0(1+\alpha_4)+[b_{1}+\alpha_5+\alpha_7]i\partial_i\gamma^i-m^3(1+\alpha_6)\right]\psi.
\end{equation}

%Performing an expansion around $d=3$ in all parameters $\alpha_i$, we find that there %is a divergence in $\alpha_5$ and $\alpha_7$, and these contributions emerge from the diagram with two vertices  and tadpole respectively. Explicitly, we have:
%\begin{eqnarray}
%\alpha_5&=&-\frac{7e^2}{12 \pi^2 (d-3) }+ \mbox{finite terms},\\
%\label{alpha6}
%\alpha_7&=&-\frac{5e^2}{4 \pi ^2(d-3) }+\mbox{finite terms}.
%\end{eqnarray}
The divergences in $\alpha_{5}$ and $\alpha_{7}$ are removed by the counterterm associated to the vertex $b_{1}$ so that $Z_{b_{1}}=1 +\delta Z_{b_{1}}$, with
\begin{equation}
 b_{1r}\delta Z_{b_{1}}=\frac{11 e^{2}}{6 \pi^{2}(d-3)}+\frac{5 e^{2}}{4 \pi^{2}}\ln \frac{M}{m}.\label{16}
 \end{equation}

Up to a term proportional to $F_{ij} \triangle^2 F_{ij}$, we may
summarize the corrections of the Maxwell and Dirac Lagrangians in the following expression: 
\begin{eqnarray}
\mathcal{L}_{IR}&=&\frac{1}{2}(1+\alpha_1)F_{0i}F_{0i}-\frac{1}{4} (a_{1r}+\alpha_{2Fin})F_{ij}F_{ij}-\frac{1}{4}(a_{2r}+\alpha_{3Fin})F_{ij}\Delta F_{ij}-\frac{a^2_3}{4}F_{ij}\Delta^2 F_{ij}+\nonumber\\ &+&
\bar\psi\left[i\partial_0\gamma^0(1+\alpha_4)+i(b_{1r}+\alpha_{5Fin}+\alpha_{7Fin})\partial_i\gamma^i-m^3(1+\alpha_6)\right]\psi,
\end{eqnarray}
where $\alpha_{5Fin}$ and $\alpha_{7Fin}$ are the finite parts of $\alpha_{5}$ and $\alpha_{7}$. In defining $\alpha_{7Fin}$ we subtract the pole term and also subtract
a term containing $\ln M/m$, as it is indicated in (\ref{16}).
\subsection{The three point vertex function}

The corrections to the three point vertex functions $\langle A_{0} \psi\bar \psi\rangle $  and $\langle A_{i} \psi\bar \psi\rangle$ receive contributions from  graphs whose structure is depicted in Fig. 3. As mentioned before, the analytic expressions for the vertex function  $\langle A_{0} \psi\bar \psi\rangle $ are finite and are associated with the  first graph,  Fig. 3(a). They are given  by
\begin{equation}
T_{10}(p,q)=-i e^3 \mu^{3-d}\int\frac{dk_{0}d^{d}k}{(2\pi)^{d+1}}\gamma_{0} S(k + p) \gamma_{0} 
  S( k + p + q) \gamma_{0}D_{00}(k),
\end{equation}
when the internal wavy line represents the propagator for the $A_{0}$ field, and
\begin{eqnarray}
T_{20}(p,q)&=&-i e^3 \mu^{3-d}\int\frac{dk_{0}d^{d}k}{(2\pi)^{d+1}}\left[ V_{3a}^{(2)}(p+k, -k) S( k + p)  
  \gamma_{0} S( k + p + q)\times \right.\nonumber\\ 
 &&\left. V_{3b}^{(2)}( p , k) D_{a b}(k)\right],
\end{eqnarray}
when the wavy line represents the propagator of the spatial component of the gauge field.
The other diagrams in Fig. 3 do not contribute to  $\langle A_{0} \psi \bar \psi\rangle $. For zero external momenta, a direct computation provides the result
\begin{eqnarray}
T_{10} &=& -\frac{i e^{3}\mu^{3-d} 2^{-d-1} \pi ^{\frac{1}{2}(-d-1)} 
   a_{3}^{\frac{(2-d)}3} m^{d-5} \Gamma
   \left(-\frac{d}{6}-\frac{7}{6}\right) \Gamma \left(\frac{d+4}{6}\right)
   \gamma_{0}}{3 \Gamma \left(\frac{d}{2}\right)},\label{t10}\\
   T_{20} &=& \frac{i e^{3}\mu^{3-d} 2^{-d-1} \pi ^{\frac{1}{2}(-d-1)} 
   a_{3}^{\frac{(2-d)}3} m^{d-5} d\Gamma
   \left(-\frac{d}{6}-\frac{7}{6}\right) \Gamma \left(\frac{d+4}{6}\right)
   \gamma_{0}}{3 \Gamma \left(\frac{d}{2}\right)}\label{t20}
\end{eqnarray}

\begin{figure}[ht]
  \begin{center}
    \vbox{
      \subfigure[]{\includegraphics[width=0.18\columnwidth]{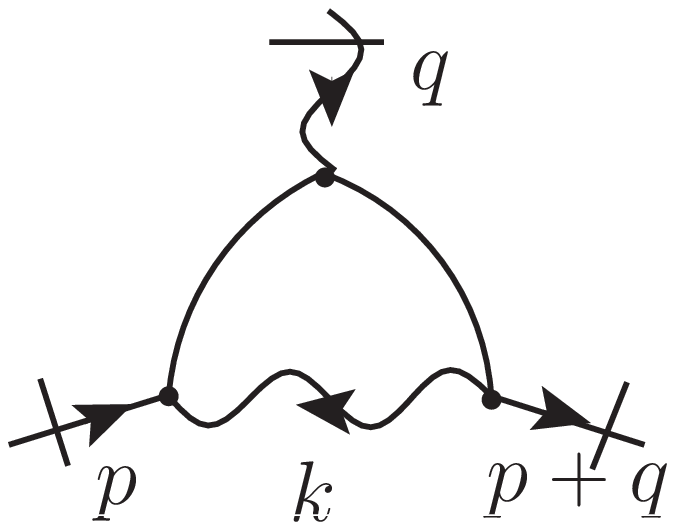}\label{fig:Figura1A}}
      \qquad
      \subfigure[]{\includegraphics[width=0.23\columnwidth]{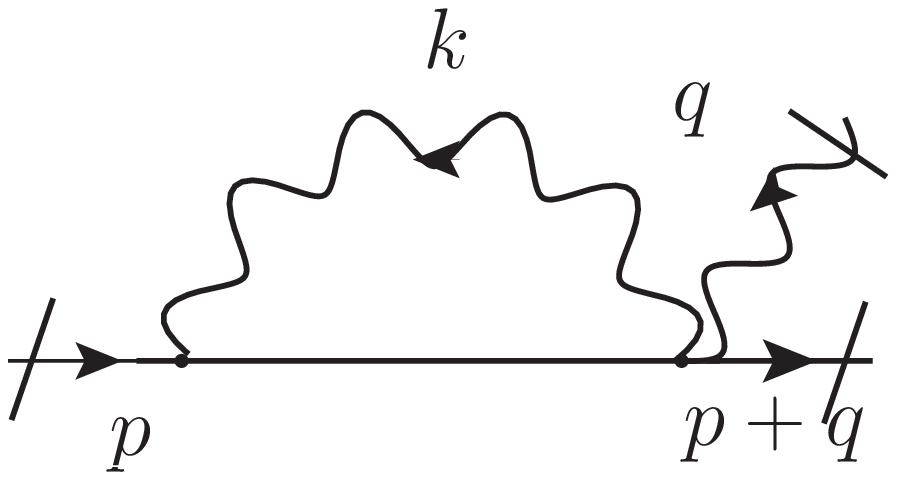}\label{fig:Figura1B}}
    \qquad
      \subfigure[]{\includegraphics[width=0.23\columnwidth]{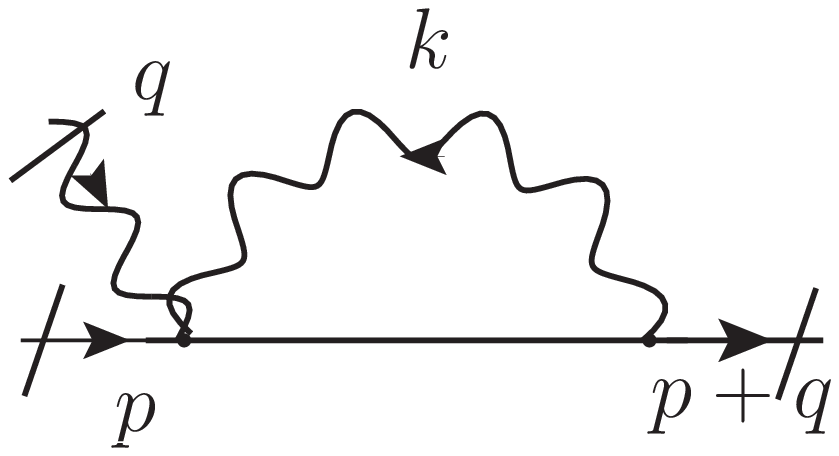}\label{fig:Figura2A}}
      \qquad
      \subfigure[]{\includegraphics[width=0.19\columnwidth]{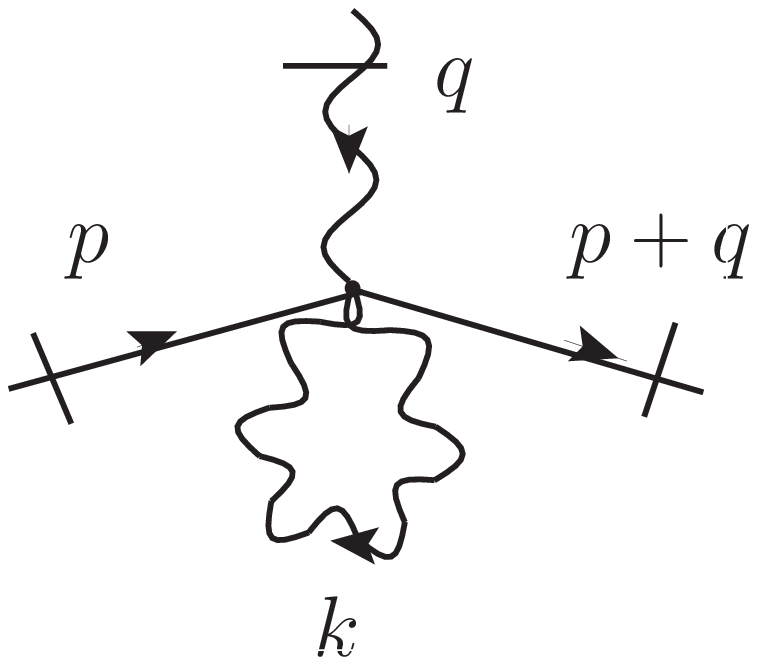}\label{fig:Figura2B}}
      }\label{Figura1}
      \caption{Graphs contributing to the three point vertex function: (a) graph with three
      trilinear vertices, graphs (b) and (c) graphs with  two vertices and (d) a tadpole graph.
      In the text, $p$ and $q$ designate the momenta entering through the fermion and gauge field lines, respectively.}
  \end{center}
\end{figure}

Concerning the vertex function $\langle A_{i} \psi\bar \psi\rangle$, we have contributions
from all graphs in Fig. 3. Thus, similarly to the previous case 
the first of those diagrams, Fig. \ref{fig:Figura1A},  is contributed by  the following expressions:

\begin{eqnarray}
T_{1i}(p,q)=-i e^2\mu^{3-d}\int\frac{dk_{0}d^{d}k}{(2\pi)^{d+1}}\gamma_{0} S(k + p)  V_{3i}^{(2)}(k + p+q, q)  
  S( k + p + q) \gamma_{0}D_{00}(k),
\end{eqnarray}
when the internal wavy line represents the propagator for the $A_{0}$ field, and
\begin{eqnarray}
T_{2i}(p,q)&=&-i\mu^{3-d}\int\frac{dk_{0}d^{d}k}{(2\pi)^{d+1}}\left[ V_{3a}^{(2)}(p+k, -k) S( k + p)  
  V_{3i}^{(2)}(k + p+q, q) S( k + p + q)\times \right.\nonumber\\ 
 &&\left. V_{3b}^{(2)}( p , k) D_{a b}(k)\right],
\end{eqnarray}
if the wavy line represents the propagator of the spatial component of the gauge field.
Besides that, there are  now also contributions with two interacting vertices, shown in 
Figs. \ref{fig:Figura1B} 
and \ref{fig:Figura2A}
\begin{equation}
T_{3i}(p,q)=-\mu^{3-d} \int\frac{dk_{0}d^{d}k}{(2\pi)^{d+1}}V_{3a}^{(2)}[(p+k,- k)  S( k + p)  
   V_{4i b} (p+q, q, k) 
  D_{a b}( -k),
\end{equation}
\begin{equation}
T_{4i}(p,q)=-\mu^{3-d} \int\frac{dk_{0}d^{d}k}{(2\pi)^{d+1}}V_{4ia} (k+p+q, q, -k) S( k + p + q) 
   V_{3b}(  p + q, k) D_{a b}( k),
\end{equation}
Finally, we have  the contribution from the tadpole graph, Fig. \ref{fig:Figura2B},
\begin{equation}
T_{5i}(p,q)=-(d + 2) a_{3} e^3 \mu^{3-d}\gamma_{i} \int\frac{dk_{0}d^{d}k}{(2\pi)^{d+1}}\frac{1}{k_0^2 - a_{3}^2 k^6 - M^6},
\end{equation}
where, to take care of infrared divergences, we again introduced the auxiliary mass parameter $M$.

Direct computation of these expressions at zero external momenta, yields
\begin{eqnarray}
\label{t1i}
T_{1i}&=&\frac{ie^3 2^{-d-1} \pi ^{-\frac{d}{2}-\frac{1}{2}}  a_3^{1-\frac{d}{3}} 
   \Gamma \left(-\frac{d}{6}-\frac{3}{2}\right) \Gamma \left(\frac{d}{6}+1\right)\left(\frac{m}{\mu}\right)^{d-3}
   \gamma_{i}}{ 9 \Gamma \left(\frac{d}{2}\right)}
  \approx-\frac{ie^{3}\gamma_{i}}{48\pi^{2}(d-3)}:\\
T_{2i}&=&\frac{ie^3 2^{-d-2} (d-2) (d (d+5)-18) \pi ^{\frac{1}{2} (-d-1)} 
   a_3^{1-\frac{d}{3}} m^{d-3} \Gamma \left(\frac{1}{2}-\frac{d}{6}\right) 
   \Gamma\left(\frac{d}{6}+1\right) \gamma_i}{3 (d+9) \Gamma \left(\frac{d}{2}+1\right)}\nonumber\\
&\approx&\frac{ie^{3}\gamma_{i}}{48\pi^{2}(d-3)};
\end{eqnarray}
\begin{eqnarray}
   T_{3i}&=& T_{4i}\nonumber\\
   &=&\frac{ie^3 2^{-d} \left(d^2-2 d+4\right) \pi ^{-\frac{d}{2}-\frac{1}{2}} 
   a_3^{1-\frac{d}{3}} m^{d-3} \Gamma \left(\frac{1}{2}-\frac{d}{6}\right) \Gamma
   \left(\frac{d}{6}+1\right) \gamma _i}{d (d+3) \Gamma \left(\frac{d}{2}\right)}\nonumber\\
   &\approx& - \frac{7ie^{3}\gamma_{i}}{24\pi^{2}(d-3)};\\
   T_{5i}&=& \frac{ie^3 2^{-d-1} (d+2) \pi ^{-\frac{d}{2}-\frac{1}{2}}  a_{3}^{1-\frac{d}{3}}
   \mu^{3-d}M^{d-3} \Gamma \left(\frac{1}{2}-\frac{d}{6}\right) \Gamma \left(\frac{d}{6}\right)
   \gamma_{i}}{3 \Gamma \left(\frac{d}{2}\right)}\nonumber\label{t5i}\\
&\approx&
   -\frac{5ie^{3}\gamma_{i}}{4\pi^{2}(d-3)} -\frac{5i e^{2}\gamma_{i}}{4\pi^{2}}\ln \frac{M}{\mu}.
   \end{eqnarray}
In the right-hand side of the above equations we explicitly quoted the pole parts at $d=3$.  Notice that the divergent terms are correctly absorbed by the renormalization of the 
parameter $b_{1}$, as it was defined in (\ref{16}).

\section{Ward identities}
In the pure gauge sector, the one loop  Ward identities  have been verified in Eqs.  (\ref{ward1}, \ref{ward2}). Here we shall consider the matter sector. In the tree approximation we have:
\begin{equation}
p_0 e \gamma^0 - p_iV_{3i}(p_1+p,-p)=ie\left(S^{-1}(p+p_1)-S^{-1}(p_1)\right).
\end{equation} 
Using this result for the two- and three-point vertex functions, in the one-loop order one should have 
\begin{eqnarray}
q^{0}T_{10}+q^iT_{1i}(p,q)&=&ie(\Sigma_1(p+q)-\Sigma_1(p));\nonumber\\
q^0 T_{20}+q^i(T_{2i}(p,q)+T_{3i}(p,q)+T_{4i}(p,q))&=&ie(\Sigma_2(p+q)-\Sigma_2(p));\nonumber\\
q^iT_{5i}&=&ie(\Sigma_3(p+q)-\Sigma_3(p)).
\end{eqnarray}
Straightforward comparison of our results for the two and three-point functions given by (\ref{sigma1}, \ref{sigma2}, \ref{sigma3}) and (\ref{t10}, \ref{t20}, \ref{t1i}-\ref{t5i}) confirms the validity of these identities up to terms linear in the momenta. 
Thus, we conclude that the Ward identities are valid not only at the tree level but also at the one-loop order.

\section{The renormalization group and Lorentz symmetry restoration}

The renormalized vertex  functions of the model, constructed by removing the pole
part of its regularized  amplitudes, depend on the scale parameter $\mu$. The investigations on the changes of this parameter  allows to relate the amplitudes at different energy scales. The implementation of the renormalization group program is greatly simplified by the fact that the divergences may be absorbed in just the parameters $a_{1}$, $a_{2}$ and $b_{1}$. Thus, because there are no wave function renormalization of the basic fields, the would be anomalous dimension is absent and the renormalization group equation for the vertex function $\Gamma^{(N_{A_{0}},N_{A_{i}},N_{\psi})}$ takes the simple form
\begin{equation}
 ( \mu \frac{\partial\phantom {a}}{\partial \mu} + \beta_{a_{1}} a_{1r}\frac{\partial\phantom a}{\partial a_{1r}}+ \beta_{a_{2}} a_{2r}\frac{\partial\phantom a}{\partial a_{2r}}+\beta_{b_{1}} b_{1r}\frac{\partial\phantom a}{\partial b_{1r}}) \Gamma^{N}[p,\kappa]=0,\label{15}
  \end{equation}  
where $N=(N_{A_{0}}, N_{A_{i}},N_{\psi})$  and the set of  parameters of the model and  external momenta are symbolically represented by $\kappa$ and $p$. The $\beta^{,}$s functions may be determined by inserting in (\ref{15}) the two point functions of  the gauge and spinor fields.  By disregarding terms of order higher than $e^{2}$, we find
\begin{equation}
\beta_{a_{1}}a_{1r}=\frac{b_{1r}e^{2}}{6 \pi^{2}},\quad \beta_{a_{2}}a_{2r}=\frac{6a_{3r}e^{2}}{5\pi^{2}},\quad \beta_{b_{1}}b_{1r}=- \frac{11 e^{2}}{6\pi^{2}}.
\end{equation}

Taking into account that $\Gamma^N(p,\kappa)$ has dimension $6-2 N_{A_{0}}-\frac{3}{2}N_{\psi}$, we may write
\begin{equation}
\bigg[ 3 p_{0}\frac{\partial}{\partial\, p_{0}}+p\frac{\partial}{\partial\, p}+
  \mu \frac{ \partial}{ \partial \mu}  
  +4 a_{1r}  \frac{ \partial}{ \partial a_{1r}} +2 a_{2r}  \frac{ \partial}{ \partial a_{2r}} 
  + 2 b_{1r}  \frac{ \partial}{ \partial b_{1r}} 
  +m  \frac{ \partial}{ \partial m} 
  - (6-2N_{A_{0}}-\frac{3}{2}N_{\psi})
  \Bigg] 
  \Gamma^{(N)} 
  = 0,
%\end{aligned}
  \label{DIMequations}
\end{equation}
so that
\begin{eqnarray}
\begin{aligned}
&\bigl [-\frac{\partial}{\partial t}+(\beta_{a_{1}}-4) a_{1r}\frac{\partial}{\partial a_{1r}}+(\beta_{a_{2}}-2) a_{2r}\frac{\partial}{\partial a_{2r}}+(\beta_{b_{1}}-2)b_{1r}\frac{\partial}{\partial b_{1r}}-e\frac{\partial}{\partial e}\bigr.\\
&-m\frac{\partial}{\partial m}\bigl.+(6-2 N_{A_{0}}-\frac{3}{2} N_{\psi})\bigr ]\Gamma^{(N)}( e^{3t}p_{0},e^{t}p,\kappa)=0.
\end{aligned} 
 \end{eqnarray} 
  At this point, we introduce effective (or  running) couplings which satisfy
\begin{equation}
\frac{\partial {\bar m}}{\partial t}=-{\bar m},\quad\frac{\partial \bar a_{1}}{\partial t}=(\beta_{\bar a_{1}}-4)\bar a_{1},\quad\frac{\partial \bar a_{2}}{\partial t}=(\beta_{\bar a_{2}}-2)\bar a_{2},\quad\frac{\partial \bar b_{1}}{\partial t}=(\beta_{\bar b_{1}}-2)\bar b_{1},\quad
\frac{\partial \bar e}{\partial t}=-\bar e, 
\end{equation}  
that are subject to the condition that at $t=0$ they are equal to the original parameters i.e., $a_{1}{(0)}=a_{1r}$, $a_{2}(0)=a_{2r}$, etc.

Notice that, as $a_{3}$ is dimensionless and does not have divergent radiative corrections,
it is independent of $t$ and may be taken very small so that, for momenta $k$ with $a_{3}k^2\ll b_{1}$,  it could be neglected in the effective Lagrangian.

 The above system may be solved by standard means furnishing  
  \begin{eqnarray}
  \bar m(t)&=& e^{-t}m, \quad \bar{e}(t)= e^{-t}e, \quad\bar{a}_{2}(t)=(a_{2r}+\frac{6a_{3r}e^{2}}{5\pi^{2}}t)e^{-2 t}, \quad \bar{b}_{1}(t)= (b_{1r}-\frac{11e^{2}}{12\pi^{2}}t) e^{-2t}\nonumber\\
{\bar a}_{1}(t) &=&[a_{1r}+\frac{b_{1r} e^{2}}{6\pi^{2}} t]e^{-4 t}.
  \end{eqnarray}

These results explicitly exhibit the asymptotic freedom of the model, all renormalized parameters vanishing at very high energies. Thus, at very high energies and momenta the perturbative methods we employed are safe. On the other hand,  for very low energies and momenta
the parameters increase and eventually these methods are no longer reliable. Actually, even the absence of Lorentz symmetry restoration, that the different behaviors of the effective parameters $\bar {a}_{1}$ and $\bar {b}_{1}$  with the energy/momenta scales seems to indicate, cannot be taken as granted.
However, before this extreme situation is reached, we may adjust $a_{3r}$ and $a_{2r}$ to very small values so that the higher derivative terms in the effective  Lagrangian may be disregarded. Thus, the model acquires a structure similar to the one described by 
\begin{equation}
{\cal L}_{1}= -\frac{1}{2} F_{i0}F_{i0}-\frac{a_{1r}}{4} F_{ij}F_{ij}+\bar\psi\left[i\gamma^0(\partial_0-ieA_0)+b_{1r}(i\gamma^i(\partial_i-ieA_i))-m\right]\psi,\label{L1}
\end{equation}
which, as argued in \cite{Taiuti1}, corresponds to QED in a material media of permittivity $\epsilon=b_{1r}$ and magnetic permeability $\mu=\frac{\epsilon}{a_{1r}}$. In \cite{Taiuti1}  the model (\ref{L1}) was attained by the clever introduction of an auxiliary "cutoff" $\Lambda_{L}$ which separates the regions of high and low energies, and letting $\Lambda_{L}$ to become very high.

\section{Summary}

We calculated the two-point functions of gauge and spinor fields and also the three point functions in a $z=3$, $d=3$ QED. After obtaining the low momenta contributions to the three point vector-spinor vertex function, we verified that our results satisfy Ward identities which confirm their validity.

 It turns out that, unlike the  $z=2$ case \cite{TM1}, in this work the two-point function of the gauge field receives a nontrivial correction, and the contribution to the two-point function of the spinor field involves the first derivatives not only in time, but also in the space sector. Compared with the $z=2$ case \cite{TM1,TM2}, this seems to indicate that the perturbative restoration of Lorentz symmetry in the low energy limit could take place.
 However, the presence of divergences precludes such a conclusion as we shortly argue. 

 The renormalization aspects of this model are quite interesting. The model is super renormalizable and up to one-loop order 
the  divergences are restricted  just to the vertices, $F_{ij}F_{ij}$, $F_{ij}\Delta F_{ij}$ and
$\bar \psi(i\gamma^i(\partial_i-ieA_i))\psi $. There are no divergent renormalization of the mass $m$ and the charge $e$ (see also \cite{nota}). 

 Furthermore, there is no wave function renormalization of the basic fields and  therefore the usual renormalization group parameters $\gamma$'s vanish. The parameter $a_{3}$, associated with the vertices with highest derivative, is dimensionless and does not receive divergent corrections. Consequently, the renormalization group equation is very simple allowing to fix  the energy dependence of the effective (running) parameters  in a straightforward way. Actually, the model is asymptotically free, all effective parameters tending to zero  for high energies/momenta.
This is a good aspect implying that at very high energies perturbative methods can be applied with confidence. On the other hand, at low energies/momenta  the effective parameters increase and the perturbative results are no longer reliable. In particular, the discussion of a possible emergence of Lorentz symmetry at low energies is hampered by this fact. The one-loop results, showing that the parameters $\bar {a}_{1}$  and $\bar{b}_{1}$ move at different rates in the energy scale, indicate that Lorentz symmetry will not hold. In that situation, the usefulness of the model will be restricted to very high energy domain where Lorentz symmetry is probably broken. 

It should be noticed  that, still being compatible with the renormalization group properties of the model, for energies which are not very low,  $a_{3}$ and $a_{2r}$ may be adjusted so that the higher derivative terms in the effective Lagrangian may be neglected. One then obtains a Lagrangian which corresponds to QED in  a material media. The occurrence of this mechanism was proposed in a earlier paper \cite{Taiuti},
in which an auxiliary cutoff $\Lambda_{L}$ separating the regions of high and low energies was used. By letting $\Lambda_{L} \to \infty$ they arrived to the mentioned effective Lagrangian. We emphasize that this outcome will hold only for energies  which are not very small.

\textbf{Acknowledgments.} The authors would like to thank the partial financial support by CNPq and CAPES. The works  of A. Yu. P. and of A. J. S.  have been partially supported by the CNPq  under the projects 303738/2015-0 and 306926/2017-2.

\begin{thebibliography}{99}
\bibitem{Kostel} V. A. Kostelecky, Phys. Rev. D69, 105009 (2004), hep-th/0312310.

\bibitem{Lifshitz} E. M. Lifshitz, Zh. Eksp. Teor. Fiz. 11, 255 \& 269 (1941).

\bibitem{Anselmi} D. Anselmi, M. Halat, Phys. Rev. D76, 125011 (2007), arXiv: 0707.2480. 

\bibitem{PH} P. Horava, Phys. Rev. D79, 084008 (2009), arXiv: 0901.3775. 

\bibitem{Visser} M. Visser, J. Phys. Conf. Ser. 314, 012022 (2012), arXiv: 1103.5587.

\bibitem{TM1} R. Iengo, M. Serone, Phys. Rev. D81, 125005 (2010), arXiv: 1003.4430;
M. Gomes, T. Mariz, J. R. Nascimento, A. Yu. Petrov, J. M. Queiruga, and A. J. da Silva, Phys. Rev. D92, 065028 (2015), arXiv:1504.04506; P. R. S. Gomes, M. Gomes, JHEP 1606, 173 (2016), arXiv: 1604.08294;
F. Marques, M. Gomes, A. J. da Silva, Phys. Rev. D96, 105023 (2017), arXiv: 1706.00796.

\bibitem{petrov1} C. F. Farias, M. Gomes, J. R. Nascimento, A. Yu. Petrov, A. J. da Silva, Phys. Rev. D85, 127701 (2012), arXiv: 1112.2081. 

\bibitem{petrov2} C. F. Farias, M. Gomes, J. R. Nascimento, A. Yu. Petrov, and A. J. da Silva, Phys.Rev. D89, 025014 (2014), arXiv: 1311.6313. 

\bibitem{petrov3} A. M. Lima, J. R. Nascimento, A. Yu. Petrov, R. F. Ribeiro,  Phys.Rev. D91, 025027 (2015), arXiv: 1412.2944. 

\bibitem{GN} A. M. Lima, T. Mariz, R. Martinez, J. R. Nascimento, A. Yu. Petrov, R. F. Ribeiro,  Phys.Rev. D95, 065031 (2017), arXiv: 1612.05900.

\bibitem{Taiuti} D. Anselmi and M. Taiuti, Phys. Rev. D {\bf 81}, 085042 (2010), [arXiv:0912.0113 [hep-ph]].

\bibitem{Bakas} I. Bakas, D. Lust, Fortsch. Phys. 59, 937 (2011), arXiv: 1103.5693; I. Bakas, Fortsch. Phys. 60, 224 (2011), arXiv: 1110.1332.

\bibitem{Dhar} A.~Dhar, G.~Mandal and S.~R.~Wadia,
  %``Asymptotically free four-fermi theory in 4 dimensions at the z=3 Lifshitz-like fixed point,''
  Phys.\ Rev.\ D {\bf 80}, 105018 (2009),
  %doi:10.1103/PhysRevD.80.105018
  arXiv:0905.2928; J.~Alexandre, J.~Brister and N.~Houston,
  %``On higher-order corrections in a four-fermion Lifshitz model,''
  Phys.\ Rev.\ D {\bf 86}, 025030 (2012),
  %doi:10.1103/PhysRevD.86.025030
  arXiv:1204.2246.

\bibitem{TM2} M. Gomes, T. Mariz, J. R. Nascimento, A. Yu. Petrov, and A. J. da Silva, Phys. Lett. B764, 277 (2017), arXiv: 1607.01240.

%\bibitem{petrov4}
%M. Gomes, J. R. Nascimento, A. Yu. Petrov, and A. J. da Silva, Phys.Rev. D90, 125022 (2014), arXiv: 1408.6499.
\bibitem{Taiuti1} D. Anselmi and M. Taiuti, Phys. Rev. D {\bf 83}, 056010 (2011), [arXiv: 1101.2019 [hep-ph]].
\bibitem{nota}  A similar model has been proposed in \cite{Taiuti}. However, the set of gauge-spinor couplings  we and \cite{Taiuti} employ  are different and consequently  our and their results are distinct.  So, unlike \cite{Taiuti} the fermion mass in our theory does not receive a divergent correction. 

\end{thebibliography}
\end{document}